# An Introduction to Lattice QCD.

Olivier Pène

Laboratoire de Physique Théorique et Hautes Energies[1]
Université de Paris XI, Bâtiment 211, 91405 Orsay Cedex, France

**Résumé.**

La méthode de calcul de QCD sur réseaux est le seul moyen de calcul non perturbatif fondé uniquement sur les bases de QCD. Après une introduction très simple aux principes de QCD sur réseaux, j'en discute les limites actuelles et la nature des processus qui lui sont accessibles. Je présente ensuite quelques résultats frappants dans les domaines des quarks légers et des quarks lourds. Enfin je tente de deviner les perspectives d'avenir.

**Abstract**

Lattice QCD is the only non-perturbative method based uniquely on the first principles of QCD. After a very simple introduction to the principles of lattice QCD, I discuss its present limitations and the type of processes it can deal with. Then I present some striking results in the light and heavy quarks sectors. Finally I try to guess the prospects.

# 1 Introduction.

Let us start with a strong statement. Lattice QCD is within our present knowledge *the non-perturbative method to solve QCD*. It needs no additional assumption beyond QCD, it has exactly as many free parameters as QCD itself, that is a coupling constant, or equivalently an energy scale ($\Lambda_{QCD}$ for example), and one mass per quark species.

However, there is a price to pay for this achievement: first, in the proper sense of a price, one needs a huge amount of computer time to reach only a limited precision. Second, only the simplest processes may be studied by this method, mainly those implying no more than one ground state hadron at a time, and in a limited region of the masses and momenta as we shall see. Still the improvements are fast, theoretical improvements and numerical ones. The lattice community is steadily stepping toward Teraflop machines, dedicated computer are used all over the world (but not in France), and we may cherish great hopes.

The principle of the method is to discretize space-time, and to work in the Lagrangian formalism, in a finite volume. Hamiltonian methods with a discretized space and a continuous time have also been tried, but they did not achieve by far the same successes and I will skip them.

Several analytic methods are available on the lattice, weak coupling (perturbative) expansions, strong coupling expansions, mean field approximations, but numerical simulations à la Monte-Carlo are by far the most fruitful, allowing to fill the gap between the strong and weak coupling regimes.

---

[1]Laboratoire associé au Centre National de la Recherche Scientifique - URA 63

Numerical simulations on lattice QCD allows you to address almost any question about confinement: you may study pure Yang-Mills QCD (i.e. QCD without quarks) and the glueball spectrum, you may study at finite temperature the deconfining phase transition as well as the chiral one, you may compute the potential between static quarks, the running coupling constant starting from non-perturbative data (and this compares quite well with direct perturbative estimates), you may learn about the QCD vacuum, its topological properties, its condensates, about the theoretical mechanism of confinement (for example you may test the dual Meissner effect picture), you may compute the parameters of the chiral Lagrangians, etc. I will here concentrate on a few topics directly related to important experimental data: computation of masses, structure functions, and electroweak matrix elements. This limited domain is still quite wide as we shall see.

In section 2. I will describe the principles of the lattice calculations, the standard actions, the use of path integrals, the Euclidean analytic continuation and the Monte-Carlo method. In section 3. I will summarize and discuss the practical limitations of the method. In section 4. I will present some striking and phenomenologically relevant results about light quarks. In section 5. I will present the same about heavy quarks. I will then conclude and try to anticipate future progress.

## 2  The principles and the computational method in short.

### 2.1  The pure gauge lattice action.

Space and time are assumed to be discretized in an hypercubic lattice. It is usual to call $a$ the lattice spacing. Let us consider all the oriented links of the lattice. A gauge field configuration is described by a set of $SU(3)$ matrices labelled by the oriented links. The relation between these $SU(3)$ matrices and the usual continuum gauge field is:

$$U_\mu(x) = P\left\{e^{iag_0 \int_0^1 d\tau A^i_\mu(x+\tau a\hat\mu)\frac{\lambda_i}{2}}\right\} \qquad (1)$$

where $U_\mu(x)$ is the $SU(3)$ matrix attached to the link starting from the site $x$ and going to the site $x + a\hat\mu$, $(\mu = 0, 3)$. $g_0$ is the bare coupling constant, $\hat\mu$ is the unity vector in the positive $\mu$ direction, $i$ is a color index and $\lambda_i$, $(i = 1, 8)$ are the so-called "Gell-Mann" Hermitean traceless $3 \times 3$ complex matrices. $P$ means a path ordered product. To the same link, oriented in the opposite direction, corresponds the inverse matrix:

$$U_{-\mu}(x + a\hat\mu) = U_\mu^{-1}(x). \qquad (2)$$

A gauge transformation is represented by an arbitrary set of $SU(3)$ matrices labelled by the sites of the lattice, $g(x)$, and acts as:

$$U_\mu(x) \to g(x)U_\mu(x)g^{-1}(x + a\hat\mu) \qquad (3)$$

A plaquette is an elementary square composed by four adjacent links, to which is attached the following $SU(3)$ matrix:

$$P(x)_{\mu,\nu} = U_\mu(x)U_\nu(x + a\hat\mu)U_\mu^{-1}(x + a\hat\nu)U_\nu^{-1}(x) \qquad (4)$$

the trace of which is gauge invariant as it is easy to check. Consequently the simplest pure gauge lattice action [1] is:

$$S[\mathcal{U}] = -\sum_{x,\mu,\nu} g_{\mu\nu} \frac{2}{g^2} \operatorname{Re} \{\operatorname{Tr}[1 - \operatorname{P(x)}_{\mu,\nu}]\}$$

$$\underset{a\to 0}{\longrightarrow} -\frac{1}{4} \sum_{i,\mu,\nu} \int d^4 x G^i_{\mu\nu}(x) G^{\mu\nu}_i(x), \quad (i = 1, 8). \tag{5}$$

for any gauge configuration $\mathcal{U}$, where a gauge configuration is defined as the set $U_\mu(x), \forall (x, \mu)$.

There exists a large number of alternative gauge invariant lattice action which have the same limit when $a \to 0$ as the action in (5). For a recent summary of the tentative improvements, see [2].

## 2.2 Path integrals.

A field theory is solved once one knows how to compute all Green functions [2]. The most elegant way to compute Green functions, which is used in lattice QCD, is through path integrals. A Green function, i.e. the T-product of several operators, is given for example by:

$$< T[\mathcal{O}(x)\mathcal{O}'(y)] > = \frac{\int \prod_{x,\mu,\nu} dU(x)_{\mu\nu} \, e^{iS[\mathcal{U}]} \mathcal{O}(x)\mathcal{O}'(y)}{\int \prod_{x,\mu,\nu} dU(x)_{\mu\nu} \, e^{iS[\mathcal{U}]}} \tag{6}$$

where $\mathcal{O}(x)$ and $\mathcal{O}'(y)$ are operators of the theory, say local polynomials of the elementary fields, $S[\mathcal{U}]$ is the action defined in (5) and $dU(x)$ is a gauge invariant integration measure on the $SU(3)$ group. The denominator insures the normalization. In (6) there is a T-product of two operators, commonly called a two-point correlation function. The generalization to n-point correlation functions is straightforward.

It is important to notice the essential simplification provided by lattice discretization and finite volume: the number of integration variables in (6) is finite, while in the continuum theory, the path integral is defined in terms of an infinite number of variables with the cardinality of the continuum! Still the integral in (6) has a huge number of integration variables: one $SU(3)$ matrix per link, i.e. in practice many millions of variables.

This makes the integral very difficult to compute, all the more so since the exponential $e^{iS[\{U\}]}$ oscillates very fast. The method used to solve the latter problem is to perform an Euclidean analytic continuation.

## 2.3 Euclidean analytic continuation.

This consists in continuing the time variable in the complex plane, and using as variables the imaginary parts: $x_4 = ix_0, p_4 = ip_0$. The Euclidean continuation of the two-point correlation function in (6) is given by:

$$\frac{\int \prod_{x,\mu,\nu} dU(x)_{\mu\nu} \, e^{-S[\mathcal{U}]} \mathcal{O}(x)\mathcal{O}'(y)}{\int \prod_{x,\mu,\nu} dU(x)_{\mu\nu} \, e^{-S[\mathcal{U}]}} \tag{7}$$

---

[2] Later on, I will give an example showing how to extract physical quantities from Green function.

where

$$S[\mathcal{U}] = \sum_{x,\mu,\nu} \frac{2}{g^2} \operatorname{Re} \left\{ \operatorname{Tr} \left[ 1 - \mathrm{P(x)}_{\mu,\nu} \right] \right\} \geq 0$$

$$\underset{a \to 0}{\longrightarrow} \frac{1}{4} \sum_{i=1,8} \int d^4x \left( G^i_{\mu\nu}(x) \right)^2 \geq 0$$

This formula allows a numerical computation as we shall see in the next subsection, thanks to the disappearance of the $e^{iS[\{U\}]}$ oscillations. Before turning to that, it is important to stress here that once having computed (7), an analytic continuation back to the Minkowskian time will be necessary. This is straightforward in the simple cases as we shall see, but for systems with more than one hadron at the same time the latter analytic continuation far off in the complex plane is practically impossible with the presently known methods. This implies that *lattice-QCD Monte-Carlo simulations are nowadays restricted to physical processes with no more than one hadron at one time*, except close to the threshold [4].

## 2.4 Monte-Carlo.

The crucial remark is that the positivity of $S[\mathcal{U}]$ in (8) gives to $e^{-S[\mathcal{U}]}$ the meaning of a probability distribution, properly normalized by the denominator in (7), and (7) is simply the formula for a mean value in a probabilistic sense. This suggests an analogy with statistical physics in four dimension which has been extensively used to reach a deeper understanding of Euclidean QCD. I will skip this and simply notice that although the probabilistic ensemble in (7) is the huge ensemble of Euclidean gauge configurations, most of them contribute for an exponentially suppressed amount. Creutz [3] suggested to use an algorithm that selects at random gauge configurations according to the probability law $e^{-S[\mathcal{U}]}/Z$ ($Z$ being the denominator in (7)). Such an algorithm will discard automatically the configurations whose contribution is negligible. Once a large number $N$ of uncorrelated configurations has been produced by the algorithm, the result is

$$\frac{\int \prod_{x,\mu,\nu} dU(x)_{\mu\nu} \, e^{-S} \mathcal{O}(x)\mathcal{O}'(y)}{\int \prod_{x,\mu,\nu} dU(x)_{\mu\nu} \, e^{-S}} = \frac{1}{N} \sum_{n=1,N} \mathcal{O}_n(x)\mathcal{O}'_n(y) + O\left(\frac{1}{\sqrt{N}}\right) \quad (8)$$

where $\mathcal{O}_n(x)$ is the operator $\mathcal{O}(x)$ with the values of the fields in the $n^{\text{th}}$ gauge configuration. Of course, many different algorithms may be used, improved, tested, and an intense activity is going on on this issue.

Although the computational task is still a formidable one, that needs hundred's of hours on supercomputers, it is already a fantastic achievement that this method has proven to be practically doable and to lead to non-perturbative predictions that agree quite well with experiment, within its rather poor numerical accuracy: $O(1/\sqrt{N})$.

## 2.5 The quark actions.

Up to now we have considered QCD with only gauge fields. What about quarks ? Quark fields are located on lattice sites, and look quite like continuum quark fields except for a different normalization: four spin and three color components. The naive action that one is tempted to use for quarks on the lattice is:

$$S_{quarks} = \sum_x \left\{ \frac{1}{2a} \left[ \overline{q}(x)(-\gamma_\mu)U_\mu(x)q(x+a\hat{\mu}) + (\hat{\mu} \to -\hat{\mu}, U_\mu \to U_\mu^{-1}) \right] + \overline{q}(x)mq(x) \right\} \quad (9)$$

whose formal limit when $a \to 0$ is the standard continuum QCD quark action and which is gauge invariant as well as chiral invariant when $m = 0$.

However, this naive action is not satisfactory because it encounters the "doubling problem". The point is that for one species of quark in (9), the quark spectrum has 16 quarks in the continuum limit $a \to 0$. I have no time to explain why this comes out, let us simply state that it has to do with the periodicity of the lattice spectrum. Many proposals have been elaborated to solve this problem [5]. Wilson's [1] proposal is simple to explain: replace the action (9) by

$$S_{quarks} = \sum_x \left\{ \frac{1}{2a} \sum_\mu \left[ \overline{q}(x)(r - \gamma_\mu)U(x)_\mu q(x + a\hat{\mu}) + (\hat{\mu} \to -\hat{\mu}, U_\mu \to U_\mu^{-1}) \right] + \overline{q}(x)(m + \frac{r}{a})q(x) \right\} \tag{10}$$

The effect of the additional terms proportional to $r$ is to yield a mass $O(r/a)$ to the 15 doublers, leaving only one quark with finite mass when $a \to 0$, as wished.

Other quark actions that improve the $a \to 0$ limit are under study [6]. It should now be clear to the reader that many lattice actions have been proposed or may be proposed for the gauge fields and many for the quark fields, leaving aside the many calculational algorithms. All these aim at the same theory, QCD. Indeed lattice QCD may be viewed as a class of QCD regularization scheme. Discretization cuts off the ultraviolet singularities. The different actions are different regularization schemes. Furthermore any regularization procedure of a field theory has to be complemented by a renormalization scheme. I will not enter here into a description of the subtle question of lattice renormalization. This is also a very ative field of research [7].

Clearly, all regularization and renormalization schemes must give the same physical quantities when $a \to 0$ and the volume goes to infinity. However, due to the numerical uncertainties and to the finite value used for $a$ and the volume, these different methods do indeed differ somehow in their physical predictions. These differences are taken as a tool to estimate the systematic errors due to finite $a$, finite volume, etc. The statistical errors due to the finite number of gauge configurations used in the Monte-Carlo may be estimated from the Monte-Carlo itself under some assumptions about the probability distribution of the quantities in question. Let us repeat, these different actions and different renormalization schemes are not at all different models, neither do they preclude to the rigor and universality of the calculations performed. The are simply different equivalent regularization/renormalization schemes for QCD.

## 2.6 An example: computation of a pseudoscalar meson mass and of $f_\pi$.

Before leaving this section, let us consider in short how one computes a physical quantity on a simple example. The axial-current two-point correlation function in Euclidean continuation verifies:

$$\int d\vec{x} < (\overline{q}(0)\gamma_0\gamma_5 q(0)) \; (\overline{q}(\vec{x}, x_4)\gamma_0\gamma_5 q(\vec{x}, x_4)) $$
$$= \sum_n | < 0|q(0)\gamma_0\gamma_5 q(0)|n > |^2 \; \frac{e^{-m_n x_4}}{2m_n}$$
$$\underset{x_4 \to \infty}{\simeq} | < 0|q(0)\gamma_0\gamma_5 q(0)|\pi > |^2 \; \frac{e^{-m_\pi x_4}}{2m_\pi}$$
$$\simeq f_\pi^2 \frac{m_\pi}{2} \; e^{-m_\pi x_4} \tag{11}$$

where the Euclidean continuation has turned the well known Minkowskian $e^{-iEx_0}$ time dependence to the $e^{-Ex_4}$ exponential decay. The sum $\sum_n$ is a sum over a complete set in the Hilbert space that is coupled to the vacuum by the axial current. At large time, thanks to the exponential decay, the lightest state dominates the sum. I call the latter state $\pi$ but it can as well be a $K$, a $D$ etc., depending on the mass of the quarks, or even, as we shall extensively use later on, a fictitious pseudoscalar meson with quark masses that do not exist in nature.

The two-point correlation function in (11) may be computed by the Monte-Carlo method described in subsection 2.4. Once the calculation performed, the large $x_4$ exponential slope gives the mass $m_\pi$, and once the latter is measured, the prefactor provides an estimate of the leptonic decay constant $f_\pi$ (renormalization introduces here a couple of subtleties which are unessential here and that I will skip). An obvious limitation of this method is that the signals produced by the excited states are difficult to extract from the statistical noise of the dominant ground state: only the ground states are reachable with a suitable statistics, except in the NRQCD that will be shortly alluded to in section 5.

## 3 The limitations.

If it is fair to say that lattice QCD is a rigorous non-perturbative method, it is also necessary to insist on its practical limitations nowadays.

### 3.1 The quenched approximation.

The quark action (10) is quadratic in the quark fields as in the continuum, and hence, as one can read in the text books, the integration of the fermionic variables in the path integral ends up in the "fermionic determinant" which is a complicated non-local effective action depending only on the gauge fields. In a simpler language, the integration of the fermionic variables amounts to compute the dynamical quark loops (i.e. the quark loops generated from the gluons) and the result is a non-local interaction between the gluons. On principle there is no obstacle in performing this calculation on the lattice and completing to the end the program described in the preceeding section. Not only is there no obstacle of principle, but it is currently done by several groups. However, these dynamical quark loops increase by several orders of magnitude the computation time and increase a lot the difficulty. And this grows even worst for larger lattices since the computation time increases like $\propto n^{4/3}$ instead of $\propto n$ for a pure gauge theory, $n$ being the number of lattice sites.

It results that for every calculation which demands a rather large number of lattice sites, a popular, although unjustified, approximation is performed: the quenched approximation. It simply amounts to replace the fermion determinant by 1, or in other words to eliminate all dynamical fermion loops such as the following one

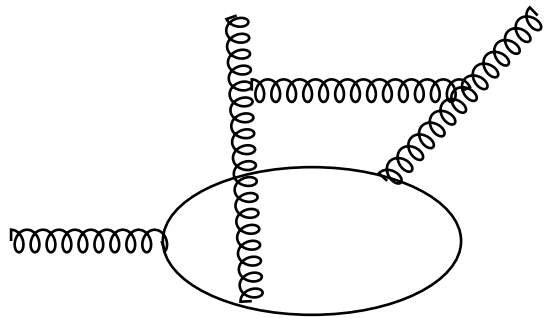

This does not mean that quarks do not propagate. In the example of section 2.6, the axial current creates/annihilates valence quarks. The latter are fully incorporated in the calculation which may be pictured as incorporating the whole set of diagrams with any number of gluon lines such as this one:

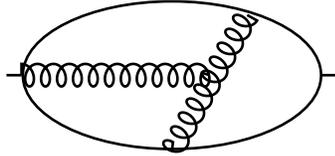

where the lines at the ends represent the insertion of the axial current.

The quenched approximation has proven to be quite satisfactory in all processes which are dominated by valence quarks. Reversely this approximation has to be avoided whenever the sea quarks dominate, and close to the chiral limit when the pion loops (the so-called "chiral loops") become important. Indeed it has been shown [8] that the chiral limit is singular in the quenched theory:

$$m_\pi^2 \propto m_q \left[1 - 2\delta \ln\left(\frac{m_\pi}{\Lambda}\right) + ...\right] \qquad (12)$$

instead of

$$m_\pi^2 \propto m_q \left[1 + O\left(m_\pi^2 \ln(m_\pi)\right) + ...\right] \qquad (13)$$

in the exact theory.

## 3.2 Limitations on masses and momenta.

To make it more concrete, the typical values of the lattice spacing practically used are $a \sim 0.05 - 0.1$ fm, i.e. $a^{-1} \sim 2 - 4$ GeV. The spatial length of the lattice is $L \sim 1 - 2$ fm; i.e. $L^{-1} \sim 100 - 200$ MeV. To reach meaningful results, the calculation has to be performed with a set of parameters such that $L^{-1} \ll m \ll a^{-1}$. Indeed, $a^{-1}$ is the ultraviolet cut-off of the lattice, and $L^{-1}$ is in some sense an infrared cut-off. A computation performed with masses of the order of $a^{-1}$ contains systematic errors that are totally out of control. Consequently, calculations may be performed with quark masses ranging from $\sim 50$ MeV ($m_\pi \sim 400$ MeV) to $\sim 2.5$ GeV. This means that only $s$ and $c$ quarks may be *directly* considered on a lattice. To study quarks $u, d, b$ some kind of extrapolation will be needed.

The momenta have also to be kept smaller than the ultraviolet cut-off. Furthermore due to finite volume and periodic boundary conditions, only a discrete set of values are possible for the momenta:

$$p_i = \frac{2\pi n}{L}, \qquad (2\pi/L \sim 0.6 - 1.\text{GeV}) \qquad (14)$$

Consequently, for light quarks, no momentum transfer larger than $5 - 10$ GeV$^2$ may be studied, and only a few discrete points are measured in the available domain.

# 4 The light quarks.

## 4.1 The chiral limit.

We have seen in the preceeding section that it is impossible to put $u$ and $d$ quarks on the presently available lattices. One then performs an extrapolation from quarks whose masses range down as close as possible to the chiral limit. In this process one uses, and simulataneously one checks, the chiral theory, i.e. the theory of spontaneous breaking of chiral symmetry.

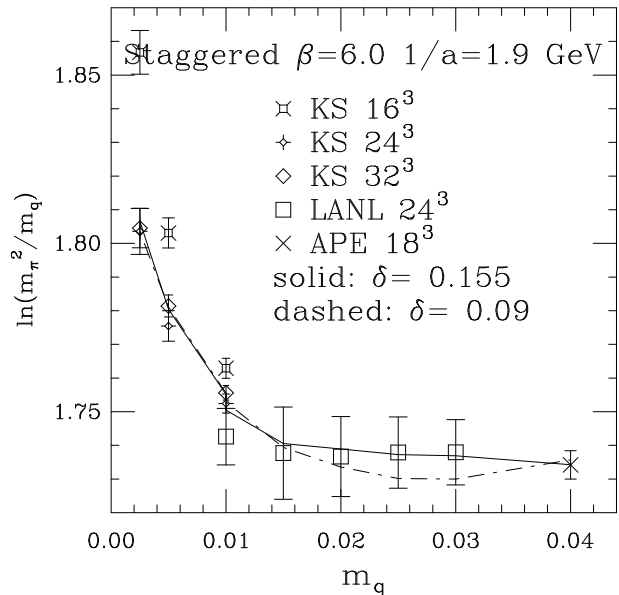

Fig. 1: The plot $\ln(m_\pi^2/m_q)$ in the quenched approximation, taken from [10] and [8], shows the $\delta$ singularity.

For example, in the case of wilson fermions (10), one *defines* massless quarks (the chiral limit) as the quarks that combine into a massless pseudoscalar meson, a Goldstone boson (the way to measure the pseudoscalar mass has been sketched in section 2.6). Notice that the parameter $m$ in eq. (10) is the quark *bare mass*. It is *not* a zero quark bare mass that corresponds to massless renormalized quarks, i.e. to a massless pseudoscalar Goldstone boson. Indeed Wilson's regularization procedure (10) breaks chiral invariance and non trivial counter-terms are needed to restore chiral symmetry in the renormalization procedure. This strategy, of using a Monte-Carlo calculation to fit the bare quark mass that corresponds to the chiral limit, is a standard example of *non-perturbative renormalization of the mass*.

A recent study of the chiral limit in the quenched approximation [9] exhibits the effect of the quenching approximation. Fig. 1 is taken from [10], [8]: it plots $m_\pi^2$ as a function of $m_q$. The points to the right of the plot show a nice flat behaviour as expected from chiral theory (13), while the points to the left show the effect of the $\delta$ singularity due to quenching (12). One learns from this plot that quenching becomes really misleading for quark masses below about one half of $m_s$. On the other hand, the linear extrapolation from masses above $m_s/2$ seems to be compatible with the chiral theory, and in fair agreement with experiment: paradoxically, within the quenched approximation, it seems better to study the chiral limit by extrapolating from not too light quarks. This restricts of course the expected accuracy.

## 4.2 $f_\pi/m_\rho$

One physical scale is necessary to renormalize lattice QCD in the chiral limit. This is necessary to fix the scale of the lattice spacing $a$. Indeed, in massless QCD there is one free scale, usually referred to as $\Lambda_{QCD}$ in perturbation theory. Hence, lattice calculations can only predict ratios of chiral quantities. For example the ratio $f_\pi/m_\rho$ (in practice $u$ and $d$ quarks can be considered as massless within the present accuracy) has been studied [11] in the quenched approximation with high statistics, and the results are reported in table 1.

| decay | finite volume | inf. volume | exp. |
|---|---|---|---|
| $f_\pi/m_\rho$ | 0.106(9) | 0.106(14) | 0.121 |
| $f_K/m_\rho$ | 0.121(6) | 0.123(9) | 0.148 |
| $F_\rho/m_\rho$ | 0.177(21) | 0.173(29) | 0.199 |
| $F_\phi/m_\rho$ | 0.217(19) | 0.253(35) | 0.219 |

Table 1: Light quarks decay constants scaled to the $\rho$ mass.

As can be seen, within errors, the quenched lattice agree with experiment (except for $f_K$) with a tendency to lie below experiment (except for $F_\phi$). Whether this tendency is a systematic quenching effect is not yet known.

### 4.3 $m_N/m_\rho$.

The $m_N/m_\rho$ is another ratio concerning light quarks that lattice compute and that can be compared with experiment. The first calculations have found a ratio around 1.4-1.5 against an experimental number of 1.20. It is now understood that this was simply due to the fact that the quark masses where too heavy in the calculations, for the reasons explained in the section 3.2.

The High Energy Monte-Carlo Grand Challenge (HEMCG) has performed an *unquenched* calculation [12], i.e. including the quark loops, of $m_N/m_\rho$ as a function of $m_\pi/m_\rho$, that we report in fig. 2. There is an indication that $m_N/m_\rho$ does indeed decrease when $m_\pi/m_\rho$ decreases, and the lattice results seem to point toward the experimental point represented on the plot by a question mark. The quenched results [13] are not significantly different from the unquenched ones, they confirm the agreement with experiment down to lower values of $m_\pi/m_\rho$.

### 4.4 Structure functions.

Structure functions are important hadronic quantities about which lattice may say a word. One usually defines the $n^{\text{th}}$ moment by:

$$\frac{2}{Q_f^2}\int dx x^n F_1(x,q^2) = \frac{1}{Q_f^2}\int dx x^{n-1} F_2(x,q^2) \equiv <x^n> \quad (15)$$

the equality being valid for $q^2 \to \infty$.

Lattice calculations have been performed for the lowest momenta some time ago [14] and seem to be confirmed by recent ones [15] which use lower quark masses. The latter find:

$$<x_u> = 0.42(4); \qquad [exp: 0.28]$$
$$<x_d> = 0.18(2); \qquad [exp: 0.11]$$
$$<x_u^2> = 0.12(2); \qquad [exp: 0.08]$$
$$<x_d^2> = 0.050(8); \qquad [exp: 0.03]$$

where [*exp*] refers to the valence quark momenta fitted by [16].

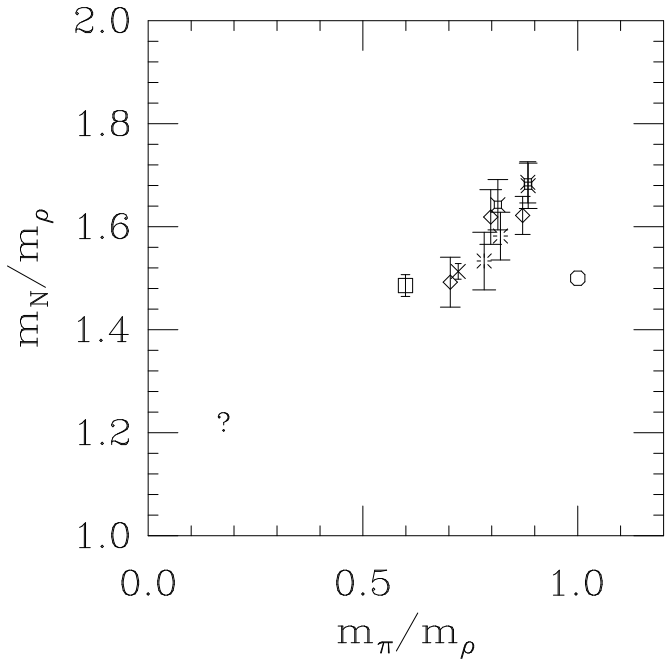

Fig. 2: $m_N/m_\rho$ as a function of $m_\pi/m_\rho$, as computed with dynamical quarks (unquenched) by the HEMCG collaboration.

It is not clear to me whether the quenching approximation may be blamed for this discrepancy.

## 5 The heavy quarks.

We have seen very few of the many lattice results about light quark physics, the preferred field for the project ELFE. Let us now turn to the realm of a $\tau$-charm factory, namely heavy flavor physics. Lattice calculations have provided a lot of results concerning masses, leptonic decay constants and semileptonic decay form factors. Notice that nonleptonic decays are out of reach because they include more than one hadron in the final state [4]. The spectroscopy of heavy quarkonia is also extensively studied including some excited states [17].

### 5.1 The charm.

Table 2, taken from [18], reports the leptonic decay constant for charmed mesons. As can be seen, the different lattice predictions agree quite well with one another as well as with experiment within the large present experimental errors. Notice the unquenched HEMCGC results, which also agree with the quenched ones within their expected larger errors. As may be seen, the lattice have an accuracy of around 10-15 % for $f_D$ and $f_{D_s}$ which were predicted at a time where no experimental data was available. In the high statistics calculations, the statistical error is as small as 5 %, the systematic errors being larger, as estimated by the authors themselves, or as indicated by the discrepancy between different groups[3].

---

[3]Different analysis methods may be used, that should agree in the continuum limit. Their difference for finite lattice spacing and finite volume gives an estimate of the systematic errors. This comparison is sometimes done inside one collaboration, but it can also be extracted from the comparison of different collaboration's results.

|        | $f_D$ (MeV)           | $f_{D_s}$ (MeV)        | $f_{D_s}/f_D$         |
|--------|-----------------------|------------------------|-----------------------|
| BLS    | $208 \pm 9 \pm 37$    | $230 \pm 7 \pm 35$     | $1.11 \pm .02 \pm .05$|
| UKQCD  | $185^{+4+42}_{-3-7}$  | $212^{+4+46}_{-4-7}$   | $1.18 \pm 0.02$       |
| PWCD   | $170 \pm 30$          |                        | $1.09 \pm .02 \pm .05$|
| APE    | $218 \pm 9$           | $240 \pm 9$            | $1.11 \pm .01$        |
| ELC    | $210 \pm 15$          | $227 \pm 15$           | $1.08 \pm .02$        |
| LANL   | $241 \pm 19$          | $266 \pm 15$           |                       |
| HEMCGC | $215 \pm 5 \pm 53$    | $288 \pm 5 \pm 63$     |                       |
| Exp.   |                       |                        |                       |
| WA75   |                       | $232 \pm 45 \pm 52$    |                       |
| CLEO   |                       | $344 \pm 37 \pm 67$    |                       |
| BES    |                       | $434^{+153+35}_{-133-33}$ |                    |

Table 2: $D_{(s)}$ leptonic decay constants from different groups, compared to experiment (in this convention: $f_\pi = 132$ MeV).

|            | $f_+(0)$          | $V(0)$            | $A_1(0)$          | $A_2(0)$          |
|------------|-------------------|-------------------|-------------------|-------------------|
| BKS        | $0.90(8)(21)$     | $1.43(45)(49)$    | $0.83(14)(28)$    | $0.59(14)(24)$    |
| UKQCD      | $0.67^{+.07}_{-.08}$ | $0.98^{+.10}_{-.13}$ | $0.70^{+.07}_{-.10}$ | $0.68^{+.11}_{-.17}$ |
| APE        | $0.72(9)$         | $1.00(20)$        | $0.64(11)$        | $0.46(34)$        |
| ELC        | $0.60(15)(7)$     | $0.86(24)$        | $0.64(16)$        | $0.40(28)(4)$     |
| LANL       | $0.73(6)$         | $1.24(8)$         | $0.66(3)$         | $0.45(19)$        |
| Exp. (average) | $0.77(4)$     | $1.16(16)$        | $0.61(5)$         | $0.45(9)$         |

Table 3: $D \to K^{(*)}$ semileptonic form factors at $q^2 = 0$ from different lattice groups, compared to experiment.

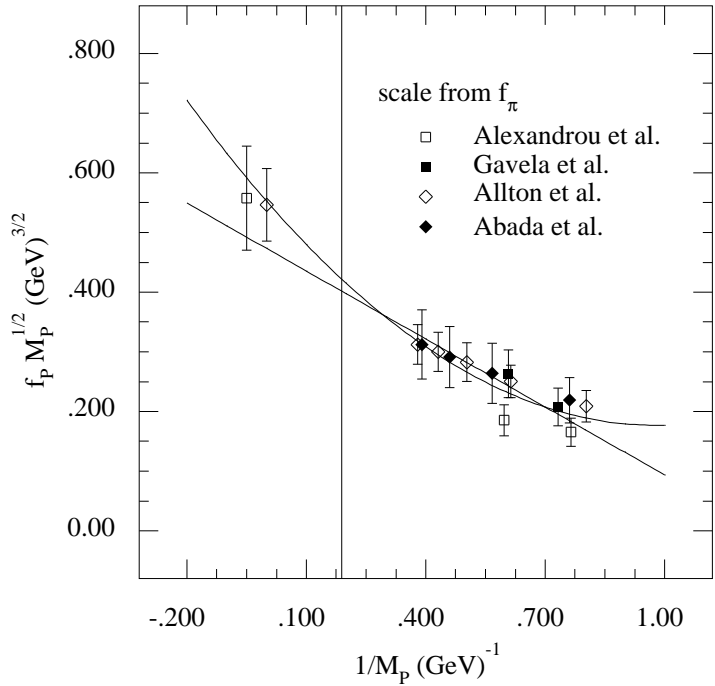

Fig. 3: $f_P M_P^{1/2}$ from [20]. The two points at the extreme left are static quarks. The points to the right correspond to moving quarks. They are in the region of the $D$ meson. All points agree. The vertical line corresponds to the $B$ meson. The slope of the fit indicates the size of non-leading, $O(1/M_P)$, corrections to HQET.

In table 3, the semileptonic form factors for $D$ decays are quoted again from [18]. The experimental average are quoted from [19]. Again the comparison with experiment is quite satisfactory for $f_+$ and $A_1$, satisfactory for $V$, while $A_2$ does not show a discrepancy within larger experimental and theoretical errors. As we have already mentioned, the finite volume lattices can only give a few points in momentum space. Practically, for $D$ semileptonic decay, in most cases only two points are in the physical $q^2$ domain. Luckily, however, the masses of $D$ and $K$ are such that these two points turn out to be at $q^2_{maximum}$ and close to $q^2 = 0$, that is they bracket the physical domain, thus limiting additional errors that would have resulted from an extrapolation.

It is now clear that lattice methods are in a good position to give their best in this domain of $D$ and $D_s$ (semi)leptonic decays. Indeed, as already stressed, $c$ and $s$ quarks are in the mass range that is easy to deal with on the lattice, and we have just seen that the $q^2$ physical domain is well accessed. It is useful to stress that a better experimental accuracy for these numbers, and particularly leptonic decay constants, from a $\tau$-charm factory or any other source, would be extremely useful to check really the precision achieved by lattice methods.

## 5.2 Extrapolation to beauty.

The $b$ quark is too heavy to be directly studied on present lattices, since $m_b > a^{-1}$. It is thus necessary to resort to an extrapolation. Similarly to the light quark case, where one uses *and* checks the chiral theory, one uses *and* checks the Heavy Quark Effective Theory (HQET) when dealing with heavy quarks. An example is provided in fig. 3, taken from [20]. HQET predicts a finite limit of $f_P M_P^{1/2}$, up to logarithms, for a heavy-light pseudoscalar meson $P$ the mass of which $M_P \to \infty$. The two points at the extreme left in fig. 3 corresponds to a direct calculation for infinite quark mass, according to a method proposed by Eichten [21] which uses *static* quarks and was one of the historical starting points of HQET. The errors on these points are rather large. The other points correspond to finite mass, *moving* quarks really computed on the lattice, i.e. to quarks in the charm region with masses ranging up to 2.5 GeV. It appears clearly from the plot that all the results are consistent, and that they show that $O(1/M_P)$ corrections to the asymptotic limit are sizeable (the slope of the straight line). The $O(1/M_P^2)$ corrections are estimated from the curvature of the parabola. A vertical line indicates the position of the $B$ meson. It may be seen that the result for $f_B$ is given by the interpolation between these two series of points. It is clear that the accuracy of the prediction for $f_B$ depends on the accuracy on $f_P$ for $M_P$ in the $M_D$ region. Hence, a theoretical improvement and experimental test of the predictions for $D$ decays is also useful to increase the accuracy of the predictions in the $B$ sector. Results from different collaborations on $f_B$, summarized in [18], are given in table 4.

| Ref. | $\beta$ | $f_B$(MeV) | $\frac{f_{B_s}}{f_{B_d}}$ |
|---|---|---|---|
| ELC(W) | 6.4 | $205 \pm 40$ | $1.08 \pm 0.06$ |
| APE (S-C) | 6.2 | $290 \pm 15 \pm 45$ | $1.11(3)$ |
| APE (S) | 6.0 | $350 \pm 40 \pm 30$ | $1.14(4)$ |
| APE (S-C) | 6.0 | $328 \pm 36$ | $1.19(5)$ |
| UKQCD (C) | 6.2 | $160^{+6}_{-6}{}^{+53}_{-19}$ | $1.22 \pm 0.04$ |
| UKQCD (S-C) | 6.2 | $253^{+16}_{-15}{}^{+105}_{-14}$ | $1.14^{+4}_{-3}$ |
| BLS (S) | 6.3 | $235(20) \pm 21$ | $1.11 \pm 0.05$ |
| Allton(S-W) | 6.0 | $310 \pm 25 \pm 50$ | $1.09 \pm 0.04$ |
| HEMCGC | 5.6 | $200 \pm 48$ | - |

Table 4: S refers to static quarks (infinite mass). The other values for $f_B$ have been interpolated to the physical $B$ mass as shown in fig. 3. It is clear in fig. 3 and in this table that the static values are higher than the interpolated ones. The HEMCGC unquenched result agrees with those obtained in the quenched approximation. For the quenched case, $\beta = 6.4$ corresponds to a lattice spacing of $\sim 0.05$ fm, while $\beta = 6.0$ corresponds to $\sim 0.1$ fm.

There exists also a specific lattice technique fit for the study of the quarkonia, and particularly the spectra of the $\Upsilon$'s (i.e. $b\bar{b}$ bound states). This is called Non Relativistic QCD (NRQCD) [22]. This method is based on a discretization of the expansion of QCD in the heavy quark velocity:

$$S_{NR} = \psi^\dagger(x) D_t \psi(x) + \psi^\dagger(x) \frac{D^2}{2M} \psi + ....  \qquad (16)$$

where $\psi$ are two-component spinor fields (quark fields are decoupled form antiquarks). This method allows also to study the first hadronic excitations, while the standard method is confined to the ground states as exemplified in section 2.6. The results seem rather good but I will skip a more detailed account of this field for lack of time. For recent reviews, see [17].

# 6  Conclusions and prospects.

Lattice QCD is a rigorous non-perturbative calculational method based on the first principles of QCD. It allows to compute numerically many simple processes implying hadrons, and to deal with a large number of phenomenological and theoretical issues.

The practical limitations in the computing power lead to the frequent use of the quenched approximation. They also yield limitations to the domain of lattice QCD: the momenta are quantized to a few allowed values due to the finite volume, and also bounded to values below or equal to $\sim 1$ GeV. Only the $s$ and $c$ quarks have masses that allow a direct study on the lattices. Generally speaking, only ground states and only systems with one hadron at a time are available.

The light quark physics is obtained through a chiral extrapolation. Quenched approximation becomes wrong very close to the chiral limit. The $b$ quark physics is obtained through an extrapolation (or interpolation) guided by the HQET.

The comparison with experiment is very encouraging within the present uncertainties which are still large: the statistical error can be as small as 5% with present statistics but the systematic error is seldom less than 15%. The unquenched calculations have still larger errors.

The state of the art is improving quite fast, thanks to a very active theoretical work, and to computer improvements. Much of the latter is due to dedicated computers built by physicists. The Gigaflop frontier has been overcome since a few years, and the present leaders are between 10 and 100 Gigaflops. The Teraflop goal is within sight. The average improvement rate is of a factor 10 every 4 years, which means a doubling of the number of lattice points in every directions every 5-6 years, or a decrease of the statistical error by a factor 4 in the same time. The theoretical work consists in improving the lattice actions to obtain a faster approach to the continuum limit ($a \to 0$), in increasing our understanding of the finite volume effects, in improving the renormalization procedure, etc. In short, the theoretical improvements aim at lessening the systematic errors. It is of course more difficult to quantify the improvements in this domain than for the statistical errors.

What accuracy will be reached in, say, ten years? I do not doubt that the statistical error will be as low as 1%, at least for the "best" physical quantities. Concerning the systematics, it is more difficult to guess. 5% ? 3% ? 1% ? One has the right to dream! Anyhow I can assure you of one thing: the lattice community *does* work hard.

# Acknowledgements

I would like to thank Asmaa Abada, Philippe Boucaud and Alain Le Yaouanc who have helped me in preparing this talk and read the manuscript. I would also like to thank the Clermont-Ferrand team who has so much contributed to the success of this meeting. This work was supported in part by the CEC Science Project SC1-CT91-0729 and by the Human Capital and Mobility Programme, contract CHRX-CT93-0132.